# Structural and magnetic properties of MnCo$_{1-x}$Fe$_x$Si alloys

J.H. Chen[a,b], Z. Y. Wei[a], E.K. Liu[a]*, X. Qi[b], W.H. Wang[a], G.H. Wu[a]

[a] *State Key Laboratory for Magnetism, Beijing National Laboratory for Condensed Matter Physics, Institute of Physics, Chinese Academy of Sciences, Beijing 100190, China*

[b] *Physics & Electronics Department, College of Science, Beijing University of Chemical Technology, Beijing 100029, China*

**Abstract:** The crystal structures, martensitic structural transitions and magnetic properties of MnCo$_{1-x}$Fe$_x$Si ($0 \leq x \leq 0.50$) alloys were studied by differential scanning calorimetry (DSC), x-ray powder diffraction (XRD) and magnetic measurements. In high-temperature paramagnetic state, the alloys undergo a martensitic structural transitions from the Ni$_2$In-type hexagonal parent phase to the TiNiSi-type orthorhombic martensite. Both the martensitic transition temperature ($T_M$) and Curie temperatures of martensite ($T_C^M$) decrease with increasing Fe content. The introduced Fe atoms establish ferromagnetic (FM) coupling between Fe-Mn atoms and destroy the double spiral antiferromagnetic (AFM) coupling in MnCoSi compound, resulting in a magnetic change in the martensite phase from a spiral AFM state to a FM state. For the alloys with $x = 0.10$, 0.15 and 0.20, a metamagnetic transition was observed in between the two magnetic states. A magnetostructural phase diagram of MnCo$_{1-x}$Fe$_x$Si ($0 \leq x \leq 0.50$) alloys was proposed.

**Keywords**: Martensitic structural transition, Metamagnetic transition, MnCoSi, Hexagonal MMX compounds

*Corresponding author.
 E-mail address: ekliu@iphy.ac.cn (E. K. Liu).

## 1. Introduction

In recent years, the magnetic hexagonal MM'X (M, M' = transition metals, X = carbon or boron group elements) alloys [1] have drawn increasing attention due to their remarkable magnetoresponsive properties, such as magnetic-field-induced martensitic transformation [2-5], giant magnetocaloric effect [3,4,6-9], magnetoresistance [10] and magneto-strain [11,12]. It is therefore of interest for the smart applications and explorations of new functional materials.

As an MM'X compound, MnCoSi alloy shows a martensitic structural transition from the Ni$_2$In-type hexagonal (*P6$_3$/mmc*, 194) austenite to the TiNiSi-type orthorhombic (*Pnma*, 62) martensite at the temperature around 1165 K [1,13]. The Néel temperature ($T_N^M$) of martensite is around 400 K [13-15]. Below $T_N^M$, the martensite phase has a special double helical magnetic structure and shows a competition between the antiferromagnetic (AFM) and ferromagnetic (FM) couplings [14,15]. According to the previous neutron diffraction studies [16], the exchange interactions between Mn-Mn and Mn-Co atoms are complicated with FM and AFM couplings coexisting in the system. Temperature [14,15], pressure [13,17], and magnetic field [14,18] can drive a first-order transition from an AFM to an FM state in MnCoSi-based compounds. Chemical composition change by substituting Ni for Co [19,20] or Ge for Si [18,21,22] in MnCoSi can also convert the AFM martensite to FM one. All these studies indicate that the helical non-collinear AFM state of MnCoSi martensite is instable and is inclined to be affected by different factors.

In this work, the MnCo$_{1-x}$Fe$_x$Si alloys with Fe atoms on the Co site were studied.

The martensitic transitions, phase structures, magnetic transitions, magnetization behavior and the origin of magnetic state conversion were analyzed. A magnetostructural phase diagram was proposed. During our studies, we noted a similar study on MnCo$_{1-x}$Fe$_x$Si alloys has been published with an emphasis on the magnetocaloric effects [23]. Our present work provides a systematical study of the basic structural and magnetic behaviors in this alloy system.

## 2. Experimental procedures

Polycrystalline ingots of MnCo$_{1-x}$Fe$_x$Si ($0 \leq x \leq 0.50$) alloys were prepared by arc-melting the metals Mn (99.99%), Co (99.99%), Fe (99.99%), and Si (99.99%) under argon atmosphere. The ingots were melted four times to obtain good alloying. All ingots were annealed for four days in evacuated quartz tubes filled with argon at 1273 K and then slowly cooled to room temperature. The X-ray diffraction (XRD) analysis with Cu-Kα was carried out with step-scan mode to characterize the crystal structures and gain the lattice parameters. The martensitic transformations were determined by differential scanning calorimetry (DSC) with a heating/cooling rate of 10 K min$^{-1}$. The magnetic measurements were performed in a superconducting quantum interference device (SQUID, Quantum Design).

## 3. Results and discussion

*3.1 Calorimetry and structure*

The thermal analysis of MnCo$_{1-x}$Fe$_x$Si alloys are performed using DSC. DSC curves are shown in Fig. 1 with the exothermic and endothermic peaks. The $T_m$ of MnCoSi ($x = 0$) is about 1131 K, which is consistent with previous reports [1,13]. For samples with $x \leq 0.50$, the $T_m$ decreases nearly linearly from 1131 K to about 823 K ($x$=0.50) with increasing $x$. This decrease may be attributed to the strengthening of the covalent bonding induced by Fe atoms in MnCo$_{1-x}$Fe$_x$Si, which is similar to the case in MnNi$_{1-x}$Fe$_x$Ge alloys analyzed by the valence-electron localization function (ELF) [4]. As the substitution level of Fe for Ni (2d sites) in MnNiGe increases, the strengthened covalent bonding improves the stability of the parent phase and consequently decreases the martensitic transformation temperature. In this study, Fe was introduced at the same Co(2d) sites, the similar strengthening of the covalent bonding and decreased $T_m$ can be obtained. It is noted that the exothermic peak becomes unobvious when the Fe content reaches $x = 0.5$, which may be caused by the Mn$_5$Si$_3$-type impurity phase.

The XRD patterns of MnCo$_{1-x}$Fe$_x$Si alloys with different Fe contents were measured at room temperature, as shown in Fig. 2(a). All the samples have a single TiNiSi-type (space group *Pnma*) orthorhombic structure. For $x = 0.50$, a coexistence of TiNiSi-type orthorhombic and Mn$_5$Si$_3$-type hexagonal structure phases is observed. From the XRD patterns, we obtained the lattice parameters of all samples and listed them in Table I. The Fe content dependence of lattice parameters is shown in Fig. 2(b). While $a_o$ shows a decreasing tendency, $b_o$, $c_o$ and the unit cell volume $V_o$ of the samples increase with increasing Fe content. It is not easy to understand these variation behaviors in this orthorhombic structure. Owing to the diffusionless and displacesive nature of martensitic transition, the relative-site occupation of atoms keeps unchanged and all atoms consistently occupy their respective sites after the transition. If we consider the relation between the orthorhombic and hexagonal parent phases: $a_o = c_h$, $b_o = a_h$, $c_o = \sqrt{3}a_h$ [1], the orthorhombic structure can be converted to the hexagonal one. In the parent phase, Mn atoms locate above and below the center of the hexatomic honeycomb Co-Si ring [1,24], forming the elemental unit in hexagonal phase, as shown in inset to Fig. 2(b). The relation between $a_h$ and the (nearest neighbor) Co-Si separation ($s_1$) is $s_1 = a_h/\sqrt{3}$. The relation between $c_h$ and the (nearest neighbor) Mn-Mn separation ($s_2$) is $s_2 = c_h/2$. In this study, the atomic radius of Fe atom (0.172 nm) is larger than that of Co atom (0.167 nm). Introducing Fe on the Co site in the Co-Si ring will increase the side length ($s_1$) of the rings. At the same time, the distance of Mn atom to the center of Co-Si ring accordingly becomes smaller due to the chemical bonding (chemical pressure) between Mn and Co/Si atoms, which results in a decrease in $s_2$. Based on the changes in $s_1$ and $s_1$, it can be seen that $a_h$ and $c_h$ will increase and decrease, respectively. Further according to the orthorhombic-hexagonal relation above, we can get that $a_o$ will decrease and $b_o$, $c_o$ increase, with increasing Fe content. Based on the orthorhombic-hexagonal structural conversion, we can well understand the variation tendency of the lattice parameters measured in this study.

*3.2 Magnetic measurements*

The temperature dependence of magnetization for MnCo$_{1-x}$Fe$_x$Si alloys measured under a magnetic field of 100 Oe is shown in Fig. 3. As shown in Fig. 3(a), the Néel temperature ($T_N^M$) of MnCoSi ($x = 0$) is 400 K, which is consistent with previous reports [13-15]. Doping a small amounts of Fe ($x = 0.10$) causes the Neel point ($T_N^M$) to change to Curie temperature ($T_C^M$ = 361 K). This indicates that this small amounts of Fe substituted on the Co site can largely change the competition balance and make the FM coupling to dominate the martensite. This behavior is very similar to the cases of MnCo$_{1-x}$Ni$_x$Si [15] and MnCoSi$_{1-x}$Ge$_x$ [21] alloys, in which doping Ni for Co or Ge for Si can also effectively destroy the non-collinear AFM state of MnCoSi and facilitates the development of FM state. The results further confirms that the exchange interactions in MnCoSi are very sensitive to the change of magnetic components and atomic distances. However, at lower temperature around 248 K, the FM state re-enters into the AFM state via metamagnetic transition, which indicates the AFM interaction in MnCoSi is still strong for this level of Fe substitution.

For alloys with 0.10 ≤ $x$ ≤ 0.30 (Figs. 3(a), 3(b) and 3(c)), the AFM-to-FM metamagnetic transitions from the high-temperature FM state to the low-temperature AFM state were observed. With increasing Fe content, the transition temperature ($T_{AFM}$) decreases from 248 K ($x$ = 0.10) to 138 K ($x$ = 0.20). The result suggests that the substitution of Fe for Co atom in present levels cannot completely overcome the AFM coupling in MnCo$_{1-x}$Fe$_x$Si alloys. This behavior is consistent with that in the same system MnCo$_{1-x}$Fe$_x$Si reported by Xu *et al*. [23], and similar to the situations in MnCo$_{1-x}$Ni$_x$Si [15] and MnCoSi$_{1-x}$Ge$_x$ [21] alloys, which also show metamagnetic transition behaviors. For higher Fe content with $x$ ≥ 0.30 (Figs. 3(d), 3(e) and 3(f), the AFM interaction at low temperatures is overcome completely and the alloys are dominated by the FM coupling. Since the exchange interaction of Fe atom is weaker than that of Co atom, $T_C^M$/$T_N^M$ of MnCo$_{1-x}$Fe$_x$Si alloys decreases from 400 K ($x$ = 0) to 296 K ($x$ = 0.40) with increasing Fe content, which is just like the case in MnCo$_{1-x}$Fe$_x$Ge alloys [25].

Figure 4(a) shows the magnetization curves for the MnCo$_{1-x}$Fe$_x$Si alloys measured at 5 K with a maximum field of 50 kOe. The magnetizing behavior varies in quite different ways, showing a FM magnetizing behavior with a high magnetization around 0.2 < $x$ < 0.40. As shown in Figure 4(b), with increasing Fe content, the saturation magnetization value ($M_s$) increases and reaches a maximum value of 3.03 μ$_B$ (120 emu/g) at $x$ = 0.30, following a decrease again when x > 0.30. The magnetic moment of alloy with $x$ = 0.25 has been calculated using the pseudopotential method with plane-wave-basis set based on the density-functional theory [26]. Although in this study there is no experimental composition at $x$ = 0.25, the corresponding calculated value of 3.47 μ$_B$ (indicated as ☆) basically falls on the experimentally measured curve of moment, showing a maximum of magnetic moment at $x$ = 0.25 in MnCo$_{1-x}$Fe$_x$Si alloys. On the contrary, $B_{cr}$ exhibits an opposite tendency and decreases slowly with increasing $x$ (0 ≤ $x$ ≤ 0.30), and then increases rapidly when $x$ > 0.30. That is, when $x$ > 0.3, $M_s$ deceases and $B_{cr}$ increases. This behavior may be related to the Mn$_5$Si$_3$-type impurity phase.

Here we try to address the origin of the conversion of the AFM to the FM state in MnCo$_{1-x}$Fe$_x$Si martensites. According to the atomic occupancy rule [14], in orthorhombic MnCoSi phase the atoms occupy three nonequivalent 4(c) sites in unit cell as follows: ($x$, 1/4, $z$), (-$x$, 3/4, -$z$), (1/2-$x$, 3/4, 1/2+$z$), (1/2+$x$, 1/4, 1/2-$z$). In MnCoSi alloy each Co atom is surrounded by six nearest-neighbor Mn atoms, forming a local Co–6Mn configuration, as shown in Fig. 5(a). The non-collinear AFM structure of MnCoSi martensite is instability and is prone to change to a canted or even collinear FM structure in a high field or under pressure [13,14,17,18]. This behavior is similar to the case of orthorhombic MnNiGe phase where the spiral AFM structure was changed to a canted FM structure in a high magnetic field [27].

Substituting of Fe for Co, Fe will occupy the Co site in MnCo$_{1-x}$Fe$_x$Si alloys (Fig. 5(b)). The chemical environment surrounding Co/Fe site keeps unchanged during the substitution except for the slight change of atomic separation. Therefore, the local Co-6Mn configuration changes into a local Fe-6Mn configuration after substitution, which is shown in Fig. 5(b). According to the magnetization behavior of MnCo$_{1-x}$Fe$_x$Si ($0 \leq x \leq 0.30$) martensites at 5 K in the field of up to 50 kOe (Fig. 4(a)), the FM coupling is established within the local Fe-6Mn configurations, which destroys the component of AFM coupling. With increasing level of Fe substitution, the regions of local Fe-6Mn FM configurations will grow up and finally overcome the native AFM interaction in the martensite phase. Correspondingly, $M_s$ rapidly increases and $B_{cr}$ decreases with $0 \leq x \leq 0.30$, which is illustrated in the inset of Fig. 4(b). In orthorhombic MnP-based magnets, it has been reported that the nearest-neighboring Mn-Mn distance d1 plays a dominant role in determining the magnetic state [11]. A smaller d1 results in an AFM sate and the larger d1 facilitates a FM state. In orthorhombic MnCoSi-based alloys, the increase in $b_o$ and $c_o$ (Fig. 2(b)) leads to an increase in d1 of MnCo$_{1-x}$Fe$_x$Si ($x \geq 0.30$) alloys and also brings about a resultant FM state, although the additional exchange interactions should be considered due to the presence of non-zero moment Co atoms. With the combined aids of the Fe-6Mn exchange interactions and moment coupling distance, the FM state is obtained in MnCo$_{1-x}$Fe$_x$Si orthorhombic martensite phase.

## 4. Conclusions

We have investigated the crystal structural and magnetic properties of MnCo$_{1-x}$Fe$_x$Si ($0 \leq x \leq 0.50$) alloys. According to our magnetic and structural measurements, we proposed a structural and magnetic phase diagram of the MnCo$_{1-x}$Fe$_x$Si system, as shown in Fig. 6. Both the Ni$_2$In-type hexagonal and TiNiSi-type orthorhombic structure are paramagnetic at high temperatures. Both the structural transition temperature $T_M$ and Curie temperature $T_c^M$ decrease with increasing Fe content up to $x = 0.50$. After Fe substitution, the FM coupling is established at the expense of the native AFM coupling, resulting in a metamagnetic transition in MnCo$_{1-x}$Fe$_x$Si ($0.10 \leq x \leq 0.20$). With increasing Fe content, the FM-AFM transition temperature $T_{AFM}$ decreases. The present work demonstrates that substituting Fe for Co atom plays an important role in converting the helical AFM structure to a FM one in the orthorhombic MnCo$_{1-x}$Fe$_x$Si martensite phase.

## Acknowledgements

This work is supported by National Natural Science Foundation of China (51301195), National Basic Research Program of China (2012CB619405), Beijing Municipal Science & Technology Commission (Z141100004214004) and Youth Innovation Promotion Association of Chinese Academy of Sciences.

Table I. Lattice constants $a_o$, $b_o$, $c_o$ axes, cell volume ($V_o$), Curie (Néel) temperature ($T_C^M/T_C^A$) and FM-AFM metamagnetic transition temperature ($T_{AFM}$) of MnCo$_{1-x}$Fe$_x$Si martensite phase. $T_M$ refers to the martensitic structural transition temperature.

| $x$ | $a_o$(Å) | $b_o$(Å) | $c_o$(Å) | $V_o$(Å$^3$) | $T_M$ (K) | $T_N^M/T_C^M$(K) | $T_{AFM}$(K) |
|---|---|---|---|---|---|---|---|
| 0 | 5.8631 | 3.6893 | 6.8595 | 148.3760 | 1162 | 400 | 400 |
| 0.10 | 5.8442 | 3.6966 | 6.8720 | 148.4584 | 1113 | 361 | 248 |
| 0.15 | 5.8342 | 3.6975 | 6.8731 | 148.2643 | 1085 | 347 | 165 |
| 0.20 | 5.8287 | 3.7017 | 6.8838 | 148.5250 | 1057 | 327 | 136 |
| 0.30 | 5.8185 | 3.7073 | 6.9061 | 148.9707 | 997 | 307 | - |
| 0.40 | 5.8005 | 3.7071 | 6.9261 | 148.9327 | 928 | 284 | - |
| 0.50 | 5.8086 | 3.7124 | 6.9405 | 149.6636 | 823 | 296 | - |

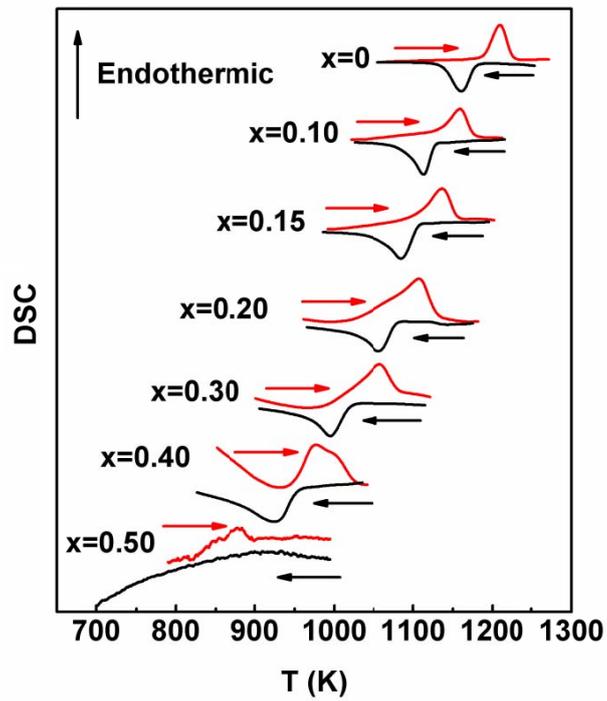

Fig. 1. DSC curves of MnCo$_{1-x}$Fe$_x$Si (0 ≤ *x* ≤ 0.50) alloys, indicating the martensitic structural transitions at high temperatures.

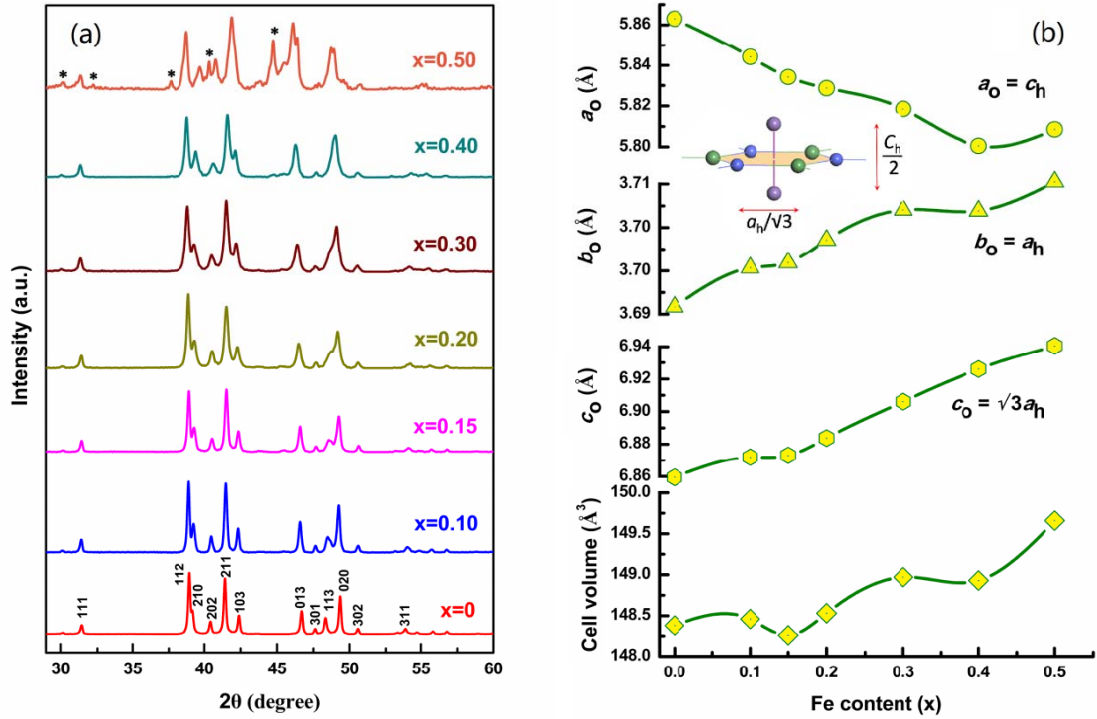

Fig. 2. Dependences of x-ray diffraction (XRD) patterns (a) and lattice parameters (b) on Fe content for MnCo$_{1-x}$Fe$_x$Si ($0 \leq x \leq 0.50$). The peaks denoted by '*' in (a) correspond to the Mn$_5$Si$_3$-type hexagonal phase. The subscripts 'h' and 'o' in (b) denote the hexagonal and the orthorhombic structures, respectively. The axes and volumes of the two structures are related as $a_o = c_h$, $b_o = a_h$, $c_o = \sqrt{3}a_h$. Inset shows the local atomic configuration with a hexatomic honeycomb ring in hexagonal parent phase.

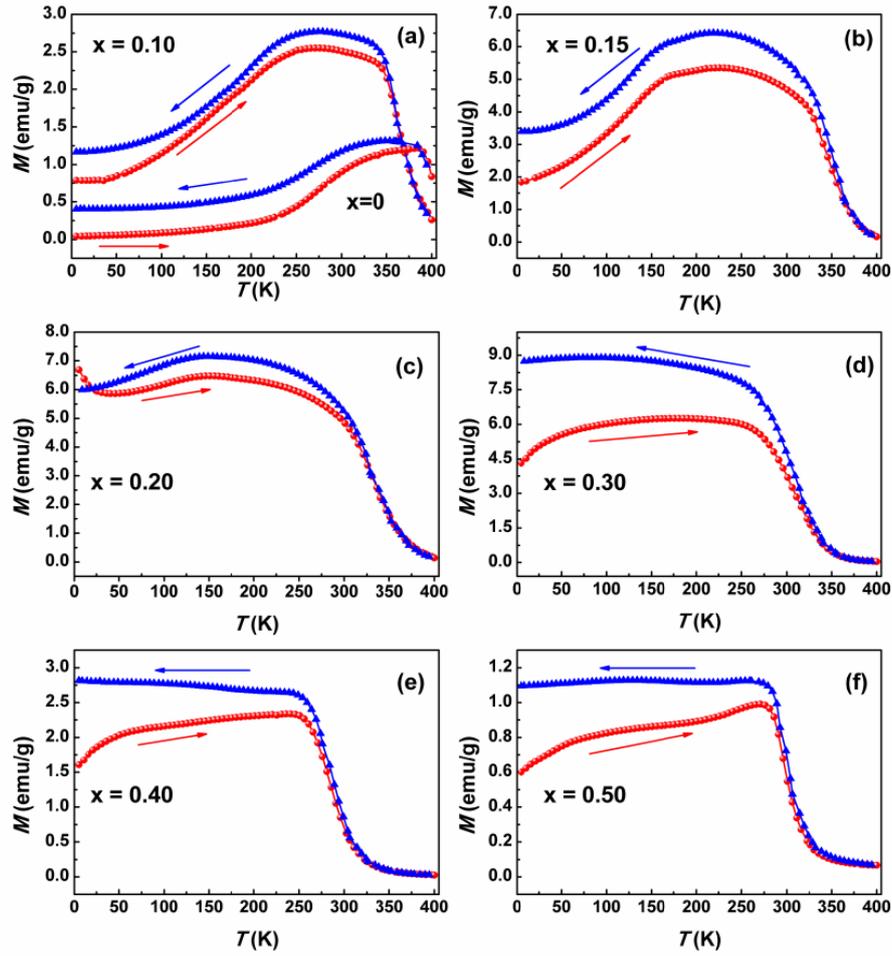

Fig. 3. Temperature dependence of magnetization of MnCo$_{1-x}$Fe$_x$Si, (a) $x = 0$ and 0.1, (b) $x = 0.15$, (c) $x = 0.20$, (d) $x = 0.30$, (e) $x = 0.40$ and (f) $x = 0.50$ in a magnetic field of 100 Oe.

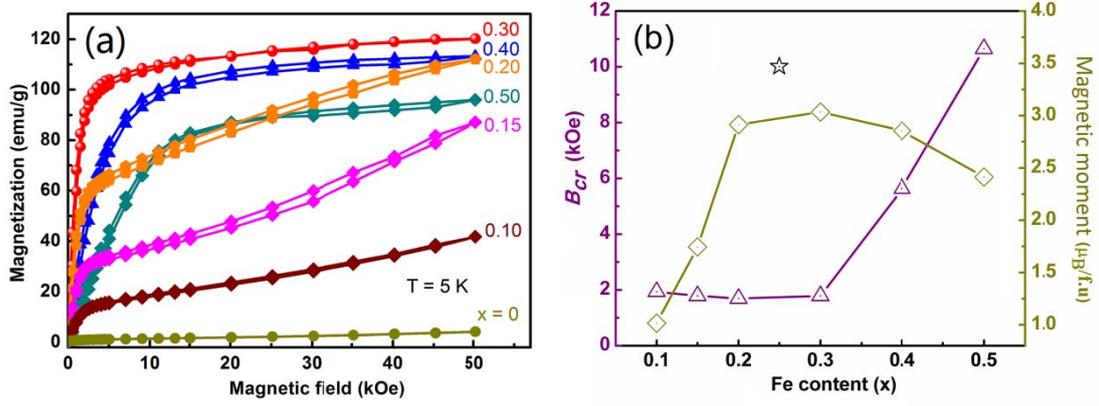

Fig. 4. (a) Magnetization curves of MnCo$_{1-x}$Fe$_x$Si ($0 \leq x \leq 0.50$) alloys measured at 5 K. (b) Fe-content dependence of $B_{cr}$ (triangle) and magnetization (diamond) in 50 kOe of MnCo$_{1-x}$Fe$_x$Si martensites. The pentagram (☆) indicates the calculated magnetic moment of FM MnCo$_{0.75}$Fe$_{0.25}$Si alloy.

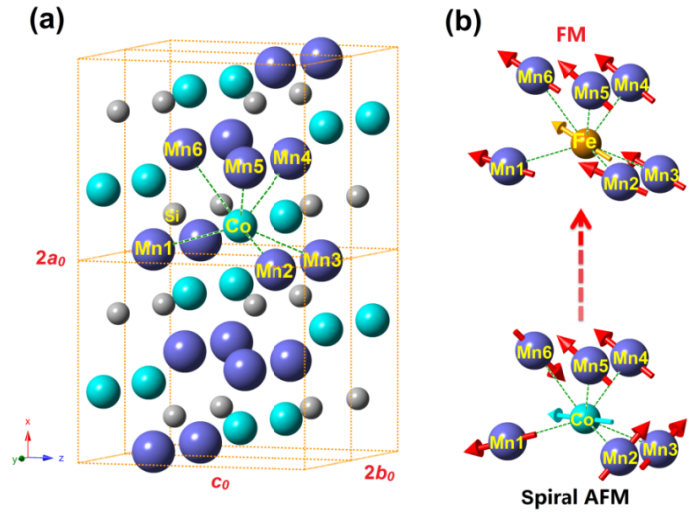

Fig. 5. (a) Crystal structure of TiNiSi-type (*Pnma*, 62) orthorhombic MnCoSi phase with indicated Co-6Mn local configurations. (b) Co-6Mn and Fe-6Mn local configurations with arrows of magnetic moment on atoms.

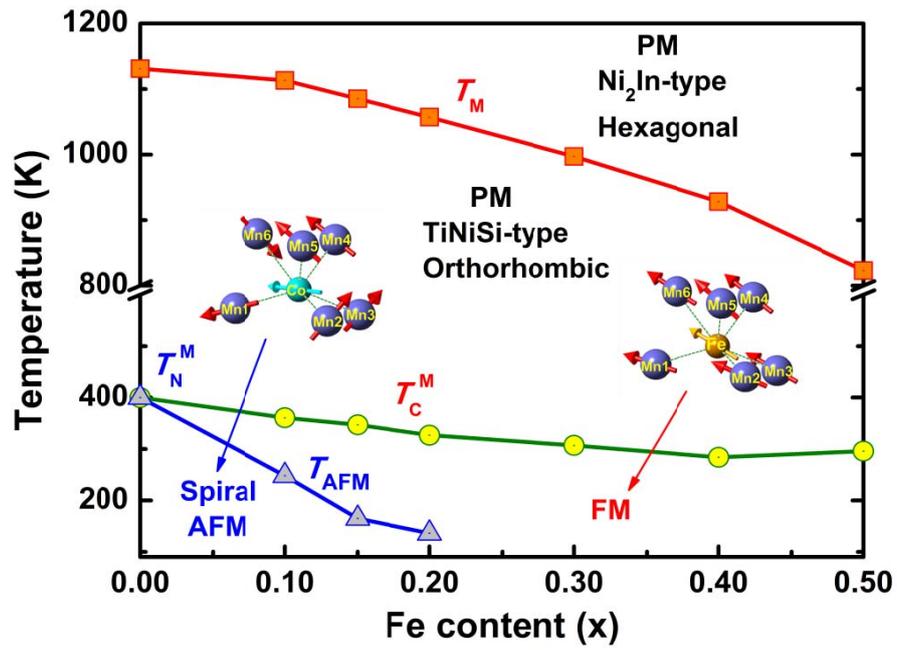

Fig. 6. Magnetostructural phase diagram of MnCo$_{1-x}$Fe$_x$Si ($0 \leq x \leq 0.50$) alloys.